\documentclass[twocolumn,8pt]{extarticle} 
\usepackage{cuted}
\usepackage{graphicx}
\usepackage{subcaption}
\usepackage{amsmath,amsfonts,amsthm}
\usepackage{cite}
\usepackage{braket}
\usepackage[a4paper, left=0.75in, right=0.75in, top=0.75in, bottom=0.8in]{geometry}
\usepackage[T1]{fontenc}
\usepackage{lmodern}
\usepackage{microtype}
\usepackage{multicol}

\renewcommand{\thesection}{\Roman{section}.}    

\title{\LARGE\bfseries
Modifying the Time-Convolutionless Master Equation
via the Moore--Penrose Pseudoinverse
}

\author{\large Caleb Blumenfeld\\[4pt]
\textit{Department of Physics and Astronomy, University of Southern California}
}

\date{\normalsize November 4, 2025}

\begin{document}

\maketitle 
\begin{strip}
\vspace*{-1.5cm}
\begin{center}
\begin{minipage}{0.7\textwidth} 
\noindent We attempt to modify the time-convolutionless master equation (TCL-ME) to be more resistant to breakdown. We remove the standard assumption that a portion of the generator is invertible by instead taking the Moore-Penrose inverse. We rederive the perturbative expansion using Israel and Charnes' result, and test the equation up to sixth and fifth orders on the Jaynes-Cummings and Ising models, respectively. We find that in both cases, the modified equation fails to capture the dynamics of the exact solution compared to the standard TCL due to the terms of the modified equation scaling exponentially with the dimension of the bath, and connect this failure to a loss of convergence of the perturbative expansion.
\end{minipage}
\end{center}
\end{strip}

\section{INTRODUCTION}
The time-convolutionless master equation (TCL-ME) of open quantum systems provides a way to accurately simulate the dynamics of quantum systems coupled to an external bath, particularly in short-time or weak-coupling cases \cite{Chen2025BenchmarkingTCL4, breuer1999,Breuer1999StochasticWaveFunction}. Because of the time-dependence of the generator $\mathcal{K}(t)$, the TCL-ME is able to simulate some non-Markovian dynamics. This allows for very accurate predictions for some systems, particularly compared to more traditional Markovian (e.g. Lindblad) approaches \cite{LarsonMavrogordatos2024,breuer1999}.

The application of the TCL-ME to quantum systems is performed via perturbative expansion of the operator $[\mathcal{I} - \Sigma(t)]$ within the generator as a power series in the perturbation parameter $\lambda$ \cite{Breuer1999StochasticWaveFunction}. However, this expansion breaks down when the operator $[\mathcal{I} - \Sigma(t)]$ becomes non-invertible. This phenomenon, known as ``TCL breakdown,'' creates a meaningful barrier to accurate simulation of certain quantum systems in the strong-coupling regime. In response, it is natural to consider using the Moore-Penrose inverse, instead of the inverse, as a way to generate a ``best-possible'' prediction of the dynamics of a system, since the former always exists and yields the least-squares solution to the associated operator equation \cite{Andersson2008}. In this work we substitute the standard inverse used in the derivation of the TCL-ME with the Moore-Penrose inverse, thereby eliminating the assumption of invertibility. 

The paper is organized as follows: In Sec. II, we give a brief review of the derivation of the TCL-ME up until the inverse of $[\mathcal{I} - \Sigma(t)]$ is taken. In Sec. III, we give a formal definition of the Moore-Penrose inverse and Israel and Charnes' power series expansion, for the latter result is not commonly known. In Sec. IV, we substitute the MP inverse into the TCL derivation and use the power series to rederive the perturbative expansion. In Sec. V, we test the new equation, dubbed the TCL+-ME for brevity, on both the Jaynes-Cummings model for a qubit in a cavity with Lorentzian spectral density and the Ising model for a qubit coupled to a bath of $N$ qubits. We choose these models because they are analytically solvable and demonstrate TCL breakdown, providing the opportunity for improvement with the TCL+-ME. We compare the predictions of the TCL+-ME with those of the TCL-ME and the analytical solution, finding no case in which the TCL+-ME offers an improvement. Finally, in Sec. VI, we provide a convergence analysis of the perturbation series to explain the failure of the TCL+-ME in the cases we tested.

\section{REVIEW OF TCL-ME DERIVATION}

A brief review of a standard derivation for the TCL-ME as in Ref. \cite{Breuer1999StochasticWaveFunction} is given for the purposes of illustrating where alteration to the power series is applied.

We begin from the von Neumann equation in the interaction picture:

\begin{equation}
\partial_{t}\rho = -i\lambda[H_I(t),\rho(t)] = 
\lambda\mathcal{L}\rho(t),
\label{vne}
\end{equation}
where $\rho(t)$ is the combined system-bath state, and the total Hamiltonian in the lab frame is 

\begin{equation}
H = H_0 + \lambda H_I,
\end{equation}
so that the interaction Hamiltonian $H_I(t)$ is $e^{iH_0t}H_Ie^{-iH_0t}$. $\lambda$ is the perturbation parameter.

\subsection{Nakajima-Zwanzig Projection Operator Technique}

We define the projector $\mathcal{P}$ as 
\begin{equation}
\mathcal{P}\rho = \operatorname{Tr}_{B}[\rho] \otimes \rho_{B}
\end{equation}
(where $\rho_B$ is a reference state of the bath) and its complement $\mathcal{Q} \equiv \mathcal{I} - \mathcal{P}$. We denote the actions of $\mathcal{P}$ and $\mathcal{Q}$ on an operator $X$ as $\hat{X}$ (``relevant part'') and $\bar{X}$ (``irrelevant part'') respectively.

The von Neumann equation \eqref{vne} for the relevant and irrelevant parts of $\rho$ read 

\begin{equation}
    \partial_{t} \hat{\rho} = \lambda\hat{\mathcal{L}}\hat{\rho} + \lambda\hat{\mathcal{L}}\bar{\rho},
    \label{relevantvne}
\end{equation}

\begin{equation}
    \partial_{t} \bar{\rho} = \lambda\bar{\mathcal{L}}\hat{\rho} + \lambda\bar{\mathcal{L}}\bar{\rho}.
    \label{irrelevantvne}
\end{equation}

\subsection{Derivation}
The general solution to \eqref{irrelevantvne} is
\begin{equation}
    \bar{\rho}(t) = \mathcal{G}_{+}(t,t_{0})\bar{\rho}(t_{0}) + \lambda\int_{t_{0}}^{t}\mathcal{G}_{+}(t,t')\bar{\mathcal{L}}(t')\hat{\rho}(t')dt',
\end{equation}
where 
\begin{equation}  
    \mathcal{G}(t,t_{0})_{+} \equiv T_{+}e^{\lambda\int_{t_{0}}^{t}\bar{\mathcal{L}}(t')dt'}
    \label{Gplus}
\end{equation},
and $T_+$ denotes the forward time-ordering superoperator. Similarly, the formal solution to the von Neumann equation \eqref{vne} is 
\begin{equation}
    \rho(t) = \mathcal{U}_{+}(t,t')\rho(t'),
\end{equation}
where 
\begin{equation}
    \mathcal{U}_{+}(t,t') \equiv T_{+}e^{\lambda\int_{t'}^{t}\mathcal{L}(s)ds}.
    \label{Uplus}
\end{equation}
$\mathcal{U}_{+}$ can then be inverted so that after applying $\mathcal{P}$ to both sides
\begin{equation}
    \hat{\rho}(t') = \hat{\mathcal{U}}_{-}(t',t)\rho,
    \label{invU}
\end{equation}
\begin{equation}
\mathcal{U}_{-}(t',t) = T_{-}e^{-\lambda\int_{t'}^{t}\mathcal{L}(t'')dt''},
\label{Uminus}
\end{equation}
where $T_{-}$ denotes backward time ordering.

Substituting \eqref{invU} into \eqref{irrelevantvne} gives
\begin{equation}
    [\mathcal{I} - \Sigma(t)]\bar{\rho}(t) = \mathcal{G}_{+}(t,t_{0})\bar{\rho}(t_{0}) + \Sigma(t)\hat{\rho}(t),
\end{equation}
where 
\begin{equation}
    \Sigma(t) \equiv \lambda\int_{t_{0}}^{t}\mathcal{G}_{+}(t,t')\bar{\mathcal{L}}(t')\hat{\mathcal{U}}_{-}(t',t)dt'.
    \label{Sigma}
\end{equation} 

It is here that the quantity $[\mathcal{I} - \Sigma(t)]$ is assumed invertible in the standard derivation of the TCL-ME. This assumption allows
\begin{align}
    \bar{\rho}(t) &= [\mathcal{I} - \Sigma(t)]^{-1}\mathcal{G}_{+}(t,t_{0})\bar{\rho}(t_{0}) \\
    &+ [\mathcal{I} - \Sigma(t)]^{-1}\Sigma(t)\hat{\rho}(t),
    \label{inverseK}
\end{align}
whereupon $\bar{\rho(t)}$ is substituted back into \eqref{relevantvne} to obtain the TCL-ME in its complete form:

\begin{equation}
    \partial_{t}\hat{\rho(t)} = \mathcal{J}(t)\rho(t_{0}) + \mathcal{K}(t)\rho(t),
\end{equation}
with
\begin{align}
    \mathcal{J}(t) &\equiv \lambda\hat{\mathcal{L}}(t)[\mathcal{I} - \Sigma(t)]^{-1}\mathcal{G}_{+}(t,t_{0})\mathcal{Q}, \\
    \mathcal{K}(t) &\equiv \lambda\hat{\mathcal{L}}(t)[\mathcal{I} - \Sigma(t)]^{-1}\mathcal{P}.
\end{align}

Under the constraint of a factorized initial state, i.e. $\rho(0) = \rho_s \otimes \rho_B$, the $\mathcal{J}(t)$ can be made to vanish.

In the case of the standard TCL, the generator $\mathcal{K}(t)$ is perturbatively expanded as 
\begin{equation}
    \mathcal{K}(t) = \lambda \hat{\mathcal{L}}(t)\sum_{k=0}^{\infty}\Sigma(t)^k,
    \label{ksum}
\end{equation}
from \eqref{sigexpansion} and $\Sigma(t)$ is further expanded as 
\begin{equation}
    \Sigma(t) = \sum_{m=0}^{\infty}\lambda^m\Sigma_m(t).
    \label{sigexpansion}
\end{equation}

\section{POWER SERIES EXPANSION FOR THE MOORE-PENROSE INVERSE}

The Moore-Penrose inverse (``pseudoinverse'') $A^{+}$ of a linear operator $A$ is a generalized inverse of A, which exists whether or not A is invertible. It is subject to the four Moore-Penrose conditions:
\begin{itemize}
    \item $AA^{+}A = A$
    \item $A^{+}AA^{+} = A^{+}$
    \item $(AA^{+})^{\dagger} = AA^{+}$
    \item $(A^{+}A)^{\dagger} = A^{+}A$
\end{itemize}

$A^{+}$ has a universal definition in the singular value decomposition:

\begin{equation}
A = UDV^{\dagger} \implies A^{+} = VD^{+}U^{\dagger},
\end{equation}
where $U$ and $V$ are unitary and $D$ is a diagonal matrix of singular values.

Additionally, for an overdetermined or square system $A\vec{x}=\vec{b}$, $\vec{x} = A^{+}\vec{b}$, represents the least-squares solution. 

Israel and Charnes \cite{benisrael1963} expand the pseudoinverse of an arbitrary square matrix with operator norm $\leq \sqrt{2}$ as
\begin{equation}
A^{+}  = \sum_{k=0}^{\infty} (\operatorname{I}- A^{\dagger}A)^{k}A^{\dagger},
\label{mpseries}
\end{equation}
which may be proven via the singular value decomposition of $A^+$ (see Appendix A).

\section{SUBSTITUTION OF MOORE-PENROSE INVERSE INTO TCL DERIVATION}

We begin the divergence from the standard TCL-ME derivation by taking the pseudoinverse of $[\mathcal{I} - \Sigma(t)]$ in \eqref{inverseK} rather than the standard inverse (indicating a least-squares or   ``best possible'' solution to $\bar{\rho}(t)$):
\begin{equation}
    \begin{aligned}
        \bar{\rho}(t) &= [\mathcal{I} - \Sigma(t)]^{+}\mathcal{G}_{+}(t,t_{0})\bar{\rho}(t_{0}) \\
        &+ [\mathcal{I} - \Sigma(t)]^{+}\Sigma(t)\hat{\rho}(t).
    \end{aligned}
    \label{inverseKmp}
\end{equation}

This eliminates the assumption that $[\mathcal{I} - \Sigma(t)]$ is invertible, as the pseudoinverse is defined for all linear operators. Naturally, we assume that a physically meaningful solution for the system exists.

With this change, we continue the derivation to find the modified $\mathcal{J}_+(t)$ and $\mathcal{K}_+(t)$ (the former of which again vanishes under factorized initial conditions):
\begin{align}
    \mathcal{J}_{mp}(t) &\equiv \lambda\hat{\mathcal{L}}(t)[\mathcal{I} - \Sigma(t)]^{+}\mathcal{G}_{+}(t,t_{0})\mathcal{Q}, \\
    \mathcal{K}_{mp}(t) &\equiv \lambda\hat{\mathcal{L}}(t)[\mathcal{I} - \Sigma(t)]^{+}\Sigma(t)\mathcal{P}.
\end{align}
\subsection{Sorting by Powers of $\lambda$}

To expand the operator $\mathcal{K}_{mp}(t)$ we apply the power series for the Moore-Penrose inverse \eqref{mpseries} instead of using \eqref{ksum}, expand the $\Sigma(t)$ terms as standard according to \eqref{sigexpansion}, and sort the resulting terms in powers of the coupling strength $\lambda$:

\begin{align}
\mathcal{K}_{mp}(t) &=\lambda \hat{\mathcal{L}}(t)\sum_{k=0}^{\infty}(\mathcal{I}-[\mathcal{I}-\Sigma(t)]^{\dagger}[\mathcal{I}-\Sigma(t)])^{k}[\mathcal{I}-\Sigma(t)]^{\dagger}\Sigma(t)\\
&= \lambda \hat{\mathcal{L}}(t)\sum_{k = 0}^{\infty}(\Sigma^{\dagger}(t) + \Sigma(t) - \Sigma^{\dagger}(t)\Sigma(t))^{k}(\mathcal{I}-\Sigma(t))^{\dagger} \Sigma(t).
\end{align}

Next, using a slight modification to the multinomial expansion theorem to account for the non-commuting operators, we conclude that the equation at order $\lambda^{n}$ will contain terms (we have dropped the time dependence notation for simplicity):

\begin{equation}
\begin{aligned}
\mathcal{K}_{mp_n}
&=
\lambda \hat{\mathcal{L}}
\sum_{p=0}^{n-2}
\sum_{\sigma\in S_n}
\sum_{\Omega_n}
\lambda^{\,ia + jb + (k+l)c + o + p}\\
&\times\mathcal{B}(a,b,c;i,j,k,l,m,n)\,
\Theta_o\,
\Sigma_p\,
\mathcal{P},
\end{aligned}
\label{uglysum}
\end{equation}

\begin{equation}
\mathcal{B}(a,b,c;i,j,k,l,m,n)
=
(\Sigma_i^a)_{\sigma_1}
(\Sigma_j^{\dagger b})_{\sigma_2}
(\Sigma_k^{\dagger m}\Sigma_l^{n})_{\sigma_3},
\end{equation}

\begin{equation}
\Theta_o
=
-\lambda^o(1-\delta_{o0})\Sigma_o
+
\delta_{o0},
\end{equation}

\begin{equation}
\Omega_n =
\left\{
\begin{aligned}
a+b+c &= p, \\
m+n &= c, \\
ia + jb + (k+l)c + o + p &= n-1
\end{aligned}
\right\},
\label{combos}
\end{equation}

\[
\sigma \in S_n,\qquad
a,b,c,o,p \in \mathbb{Z}_{\ge 0},\qquad
i,j,k,l,m,n \in \mathbb{Z}_{>0}.
\]

\textit{**A note: the above is why the standard choice of $\Sigma$ for the expansion operators is asinine}.\\

Where the sum $\sum_{p}$ is over powers, as in the multinomial theorem, $\sum_{\sigma \in S{n}}$ is over all permutations of the $\sigma_{n}$ terms in $\mathcal{B}$, and $\sum_{\Omega_n}$ is over the  index combinations defined in \eqref{combos}.

Matching powers of $\lambda$ gives, for the first 4 terms:

\begin{equation}
    \begin{aligned}
    \lambda_{1} & : 0 \\
    \lambda_{2} & : \Sigma_{1} \\
    \lambda_{3} & : \Sigma_{1}^{2} + \Sigma_{2} \\
    \lambda_{4} & : \Sigma_{1}^{3} + \Sigma_{1}\Sigma_{2} + \Sigma_{2}\Sigma_{1} 
                  + \Sigma_{3} + 2\Sigma_{1}^{\dagger}\Sigma_{2}.
    \end{aligned}
\end{equation}

In comparison, the first 4 terms for the regular TCL-ME expansion taken from the inverse series read:

\noindent
\begin{equation}
\begin{aligned}
\lambda_{1} & : 0\\
\lambda_{2} & : \Sigma_{1} \\
\lambda_{3} & : \Sigma_{1}^{2} + \Sigma_{2} \\
\lambda_{4} & : \Sigma_{1}^{3} + \Sigma_{1}\Sigma_{2} + \Sigma_{2}\Sigma_{1} + \Sigma_{3}.
\end{aligned}
\end{equation}

Using the pseudoinverse series up to 4th order results in an identical expansion to the inverse series, with one extra term: $2 \Sigma_{1}^{\dagger} \Sigma_{1}^2$.

Further, because omitting all $\Sigma^{\dagger}$ contributions from the pseudoinverse series yields the regular Neumann series, we note that the TCL+-ME expansion is identical to the standard TCL-ME structure with extra $\Sigma^{\dagger}$ terms added at each order. This simplifies the construction of the TCL+: one only needs to compute the additional $\Sigma^{\dagger}$ contributions and add them to the standard TCL terms, which can be efficiently obtained via ordered cumulants (see Sec. V A).

\subsection{Adjoint Operators}
The form of the adjoint $\Sigma$ operators of the TCL+ expansion are proven in this section. 

By expanding $\mathcal{G}(t,t')$ \eqref{Gplus} and $\mathcal{U_-}(t',t)$ \eqref{Uminus} into powers of $\lambda$, we may obtain the $\Sigma_m(t)$ terms from \eqref{sigexpansion}. The first two terms are written out explicitly below:

\begin{equation}
    \Sigma_{1}(t) = \int_{t_{0}}^{t}\mathcal{L}(t')\mathcal{P}dt'
\end{equation},
\begin{equation}
\begin{aligned}
    \Sigma_{2}(t) &= \int_{t_{0}}^{t}dt''\int_{t_{0}}^{t''}dt'[\mathcal{L}(t'')\mathcal{L}(t')\mathcal{P} \\
    &- \mathcal{P}\mathcal{L}(t'')\mathcal{L}(t')\mathcal{P} - \mathcal{L}(t')\mathcal{P}\mathcal{L}(t'')].
\end{aligned}
\end{equation}

By linearity, the adjoints of these operators are obtained  via reordering the operators inside the integrals:
\begin{equation}
\Sigma_{1}^{\dagger}(t) = \int_{t_{0}}^{t}\mathcal{P}^{\dagger}\mathcal{L}^{\dagger}(t')dt',
\end{equation}
\begin{equation}
\begin{aligned}
    \Sigma_{2}^{\dagger}(t)&=\int_{t_{0}}^{t}dt''\int_{t_{0}}^{t''}dt'[\mathcal{P}^{\dagger}\mathcal{L}^{\dagger}(t')\mathcal{L}^{\dagger}(t'')\\
    &-\mathcal{P}^{\dagger}\mathcal{L}^{\dagger}(t')\mathcal{L}^{\dagger}(t'')\mathcal{P}^{\dagger} - 
    \mathcal{L}^{\dagger}(t'')\mathcal{P}^{\dagger}\mathcal{L}^{\dagger}(t')].
\end{aligned}
\end{equation}

The adjoints of the $\mathcal{P}$ and $\mathcal{L}$ operators are shown to be the following:

\begin{equation}
    \mathcal{L}^{\dagger} = i[H(t),\cdot]
    \label{Ldag},
\end{equation}

\begin{equation}
    \mathcal{P}^{\dagger} = \operatorname{Tr}_{B}[\cdot (I_{s} \otimes \rho_{b})] \otimes I_{b}.
    \label{Pdag}
\end{equation}

A proof of the validity of these adjoints is provided in Appendix B.

\subsection{$\mathcal{P}\mathcal{L}\mathcal{P} = 0$ Relations for Adjoints}

One of the most useful ways of simplifying the TCL-ME is by choosing a modified basis operators for the bath in $H(t)$ to impose the relation $\mathcal{P}\mathcal{L}\mathcal{P} = 0$, which allows for the cancellation of many terms \cite{breuer1999, Breuer1999StochasticWaveFunction}. In analogy to this, we develop similar relations to include the adjoints of the $\mathcal{P}$ and $\mathcal{L}$ operators. First

\begin{equation}
    \mathcal{P}^{\dagger}\mathcal{L}^{\dagger}\mathcal{P}^{\dagger}X = 0 \quad \forall X \in \mathcal{B}(\mathcal{H}_{S} \otimes \mathcal{H}_{B})
\end{equation}
is trivial, as it is equivalent to $0^{\dagger} = 0$. Of more interest, however, we also find

\begin{equation}
    \mathcal{P}\mathcal{L}\mathcal{P}^{\dagger} \neq 0,
\end{equation}

\begin{equation}
    \mathcal{P}^{\dagger}\mathcal{L}\mathcal{P} \neq 0,
\end{equation}
in general. To see this, we represent $\mathcal{H}_{SB}$ as $\sum_{\alpha} A_{\alpha} \otimes B_{\alpha}$. Then:

\begin{equation}
\begin{aligned}
    \mathcal{P}^{\dagger}\mathcal{L}\mathcal{P}X 
    &= -i\sum_{\alpha}\operatorname{Tr}_{B}\!\bigg[
    [A_{\alpha} \otimes B_{\alpha},\\ &\operatorname{Tr}_{B}[X]\otimes\rho_B](I_{S}\otimes\rho_{B})\bigg] \otimes I_{B},
\end{aligned}
\end{equation}
which simplifies to 

\begin{equation}
    -i\sum_{\alpha}[A_{\alpha}, \operatorname{Tr}_{B}[X]]\operatorname{Tr}[B_{\alpha}\rho_B^2]\otimes I_{B}.
\end{equation}

 In the standard derivation, we force  $\mathcal{P}\mathcal{L}\mathcal{P} = 0$ by setting $\operatorname{Tr}[B_{\alpha}\rho_{B}] = 0$ through a shift in the bath operators: 
 \begin{equation}
     B' = B - \langle B \rangle I_{B}
 \end{equation} where 
 \begin{equation}
     \langle B \rangle \equiv \operatorname{Tr}[\rho_{B}B]
 \end{equation}
 is the expectation value of B.
 
 However, for $\mathcal{P}^{\dagger}\mathcal{L}\mathcal{P}$, this term is replaced by $\operatorname{Tr}[B_{\alpha}\rho_B^2]$, which is nonzero in general (barring $\rho_B$ pure). Similarly, for $\mathcal{P}\mathcal{L}\mathcal{P}^{\dagger}$ we obtain $\operatorname{Tr}[B_{\alpha}]$, again which again need not be 0. Therefore we cannot in general simplify the TCL+-ME unless the reference state is idempotent or the bath operators are traceless.

\section{TESTING}
In this section we apply the TCL+-ME to two analytically solvable models in order to compare its predictions with those of the TCL-ME. First, we examine the Jaynes-Cummings model of a qubit in a harmonic oscillator bath (oft used to compare the behavior of master equations \cite{Chen2025BenchmarkingTCL4,LarsonMavrogordatos2024,Breuer1999StochasticWaveFunction, Lampert2025}). Next, we test the equation on the Ising model for a qubit in a bath of other qubits.

\subsection{Jaynes-Cummings Model}
The Jaynes-Cummings Hamiltonian in the interaction picture is given by 

 \begin{equation}
     H(t) = \sigma_+(t) \otimes B(t) + \sigma_- \otimes B^\dagger,
 \end{equation}

 \begin{equation}
     \sigma_+(t) = e^{i\Omega_0t}\sigma_+,
     B(t) = \sum_ke^{-i\omega_kt}g_kb_k,
 \end{equation}
where $\sigma_- = \ket{0}\bra{1} = \sigma_+^\dagger$ is the two-level lowering operator of the system, the $b_k$ are the bosonic lowering operators for bath mode $k$, and $\Omega_0$ and $\omega_k$ are the energy of the system excited state and bath mode k, respectively.

We choose the reference state as the ground state of the oscillator,

\begin{equation}
    \rho_B = \ket{0}\bra{0}_B.
\end{equation}

\begin{figure*}[t]  
\centering
\begin{subfigure}[b]{0.32\textwidth}
    \centering
    \includegraphics[width=\linewidth]{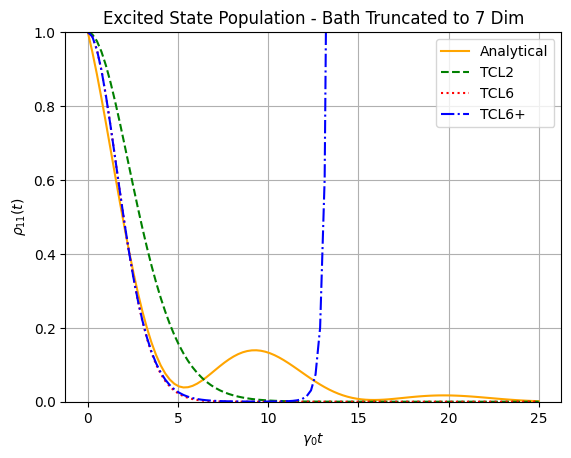}
    \caption{1 Dim}
\end{subfigure}
\hfill
\begin{subfigure}[b]{0.32\textwidth}
    \centering
    \includegraphics[width=\linewidth]{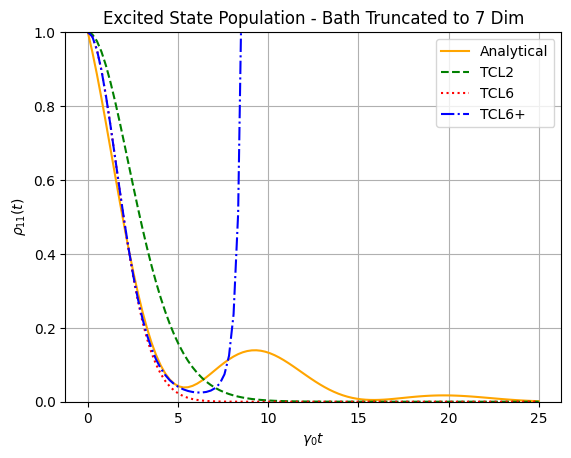}
    \caption{3 Dims}
\end{subfigure}
\hfill
\begin{subfigure}[b]{0.32\textwidth}
    \centering
    \includegraphics[width=\linewidth]{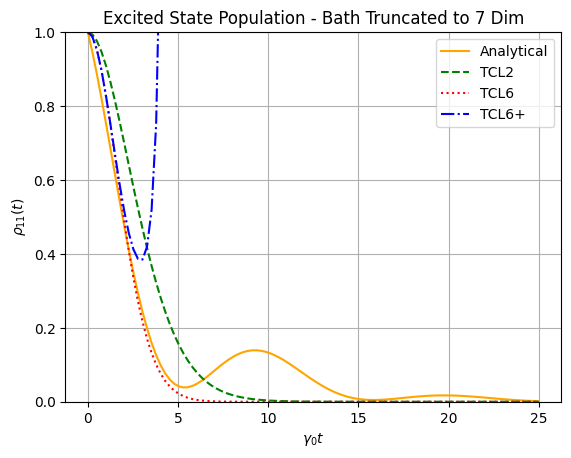}
    \caption{7 Dims}
\end{subfigure}
\caption{Comparison of the analytical solution, TCL-2, TCL-6, and TCL+-6 with a truncated bath dimension in the Jaynes-Cummings model. Note the divergent behavior of the TCL+ as the bath dimension increases. }
\label{fig:jcplots}
\end{figure*}

\begin{figure*}[t]  
\centering
\begin{subfigure}[b]{0.45\textwidth}
    \centering
    \includegraphics[width=\linewidth]{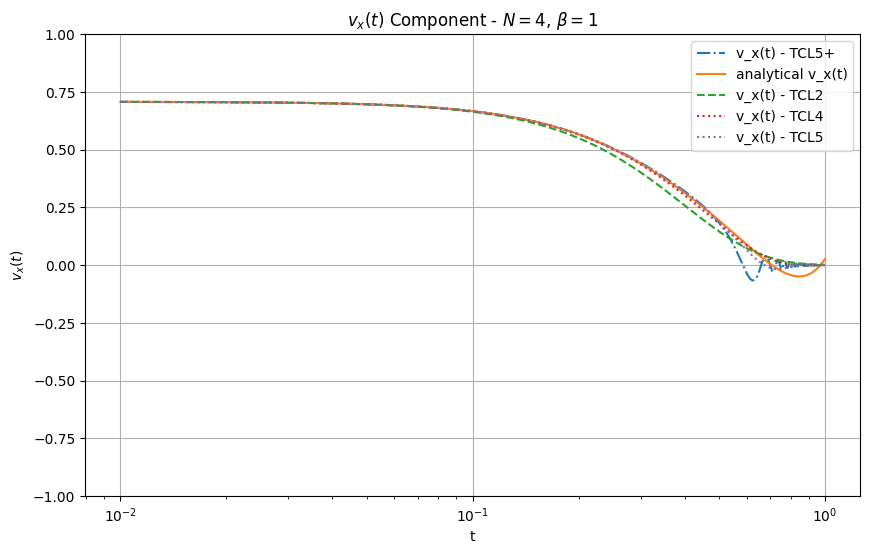}
    \caption{4 bath qubits, $\beta = 1$}
\end{subfigure}
\hfill
\begin{subfigure}[b]{0.45\textwidth}
    \centering
    \includegraphics[width=\linewidth]{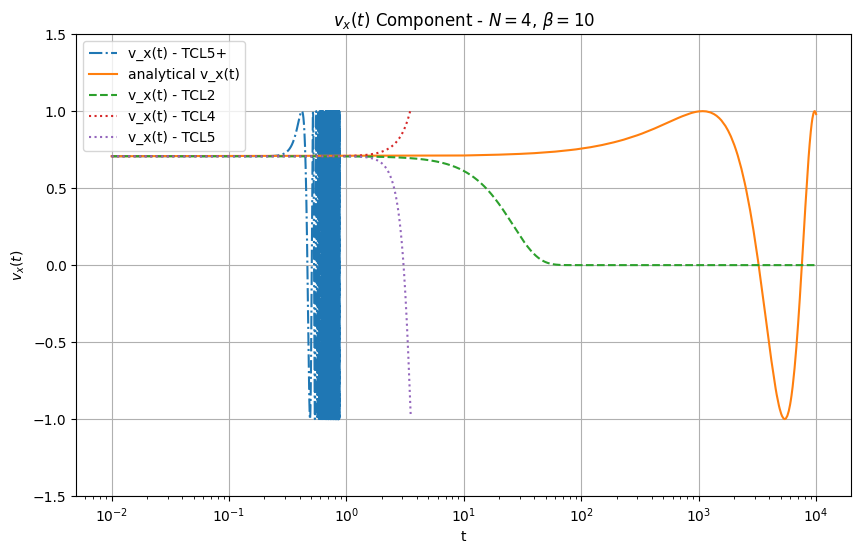}
    \caption{4 bath qubits, $\beta = 10$}
\end{subfigure}

\vspace{1em} 

\begin{subfigure}[b]{0.45\textwidth}
    \centering
    \includegraphics[width=\linewidth]{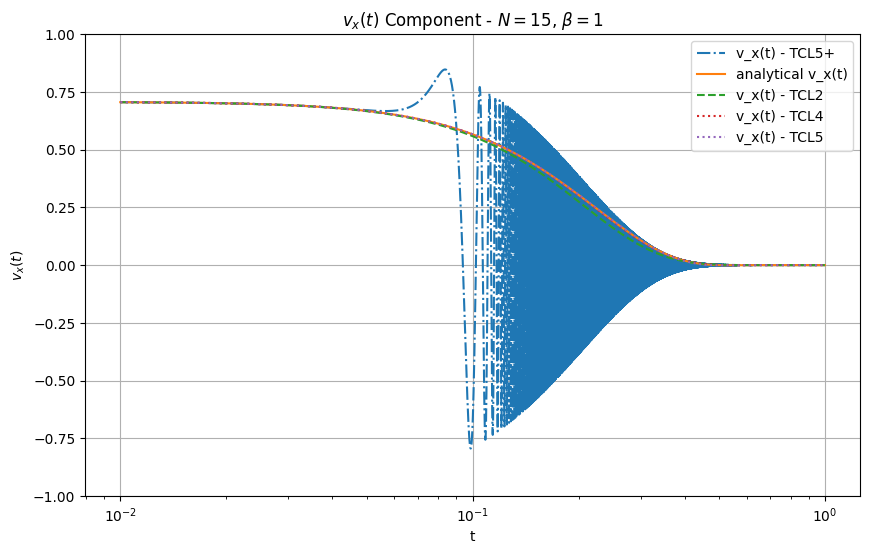}
    \caption{15 bath qubits, $\beta = 1$}
\end{subfigure}
\hfill
\begin{subfigure}[b]{0.45\textwidth}
    \centering
    \includegraphics[width=\linewidth]{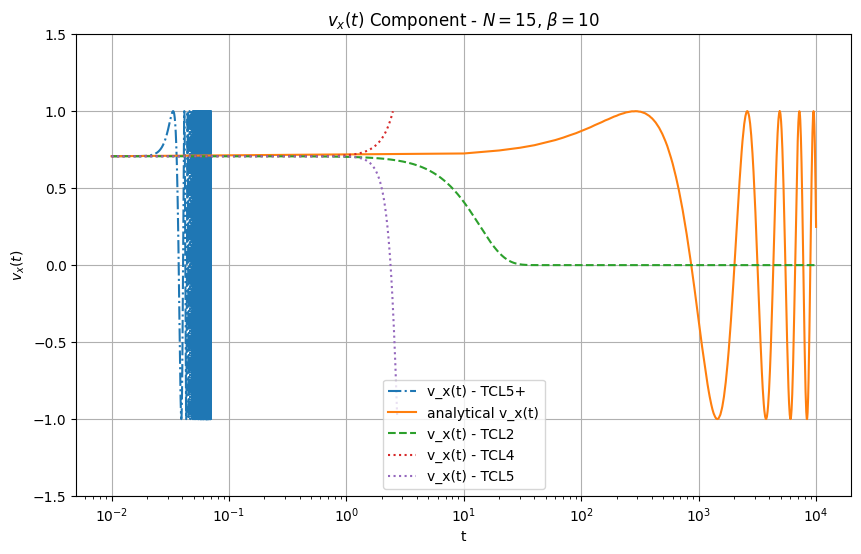}
    \caption{15 bath qubits, $\beta = 10$}
\end{subfigure}

\caption{Simulation of the analytical solution, TCL-2,4,5, and TCL+-5 for all combinations of $N = (4,15)$ and $\beta = (1,10)$ in the Ising Model. Because the only extra term is at 5th order, we only see divergence of the TCL+ in frequency, not amplitude.}
\label{fig:jcplots}
\end{figure*}

For the TCL terms, we can extend the use of the $\mathcal{P}\mathcal{L}\mathcal{P} = 0$ relations to cancel out all terms with an odd number of $\mathcal{L}$s between $\mathcal{P}$s due to the Gaussianity of the bath. We recognize that the structure of all terms in the standard TCL, then, will be only even numbers of $\mathcal{L}$s between $\mathcal{P}$s.

The Jaynes-Cummings model can be solved analytically for the 1-excitation subspace, in which we make simplifying assumption that the system can support no more than one excitation \cite{breuer1999, Breuer1999StochasticWaveFunction}. Formally, we express the possible states the system can take as 

\begin{equation}
    \ket{\psi_0} = \ket{0}_S \otimes \ket{0}_B,
\end{equation}

\begin{equation}
    \ket{\psi_1} = \ket{1}_S \otimes \ket{0}_B,
\end{equation}

\begin{equation}
    \ket{\phi_k} = \ket{0}_S \otimes \ket{k}_B,
\end{equation}
where $\ket{k}_B$ denotes the bath state with an excitation in mode k. The joint system-bath state is 

\begin{equation}
    \ket{\phi(t)} = c_0(t)\ket{\psi_0} + c_1(t)\ket{\psi_1} + \sum_kc_k(t)\ket{\phi_k}.
\end{equation}

The analytical solution to the JC model in the 1-excitation subspace is treated in several different works \cite{breuer1999,Breuer1999StochasticWaveFunction}, so we will simply state the resulting system density matrix $\rho_s(t)$ below:

\begin{equation}
    \rho_s(t) = \begin{bmatrix}
1 - |c_1|^{2}  & c_0c_1^*(t) \\
c_0^*c_1(t) & |c_1|^2
\end{bmatrix}
\end{equation}
($c_0$ is found to be constant in time).

The exact master equation for the system is 
\begin{equation}
\begin{aligned}
    K_S(t)\rho(t) &= -\frac{i}{2} S(t)\,[\sigma_{+} \sigma_{-}, \rho(t)] \\
    &+ \gamma(t)\left(\sigma_{-} \rho(t) \sigma_{+} - \tfrac{1}{2}\{\sigma_{+} \sigma_{-}, \rho(t)\}\right).
\end{aligned}
\end{equation}
which is conveniently in TCL form, with our reduced generator $\mathcal{K}_S(t)$ for the system related to the regular generator $\mathcal{K}(t)$ by $\mathcal{K}_s(t) = \operatorname{Tr}_B[\mathcal{K}(t)(\rho_s(t) \otimes \rho_B)]$. We then recognize that $\sigma_+$ is an eigenoperator of $\mathcal{K}_S(t)$:

\begin{equation}
    K_s(t)\sigma_+(t) = -\frac{1}{2}[\gamma(t) + iS(t)]\sigma_+.
    \label{pluseigen}
\end{equation}

We subsequently find that $\sigma_+ \otimes \rho_B$ is also an eigenoperator each individual expanded term of the TCL+. In this way, the TCL+ is analogous to an expansion of the rates $\gamma(t)$ and $S(t)$ in powers of $\lambda$. 

Recalling the structure of the TCL+-ME, we may divide the work of applying $\sigma_+ \otimes \rho_B$ to to the TCL terms and the adjoint terms respectively. 

\begin{equation}
\begin{aligned}
    \mathcal{K}_S(t)\sigma_{+}
&= \sum_{n=1}^{\infty} \lambda^{2n}
\int_{0}^{t} dt_{1} \int_{0}^{t_{1}} dt_{2} \cdots \int_{0}^{t_{2n-2}} dt_{2n-1} \,\\
&\times\operatorname{Tr}_{B}\!\left[\langle \mathcal{L}(t)\mathcal{L}(t_{1}) \cdots \mathcal{L}(t_{2n-1}) \rangle_{OC} \sigma_{+} \otimes \rho_{B} \right].
\end{aligned}
\end{equation}
where the $\langle \cdots \rangle_{OC}$ denotes ordered cumulants of the terms and are defined in Refs. \cite{ShibataArimitsu1980,breuer1999}.

We then make use of two relations:

\begin{equation}
    \mathcal{P}\mathcal{L}(t)\mathcal{L}(t_1)\sigma_+ \otimes \rho_B = f(t-t_1)\sigma_+ \otimes \rho_B,
    \label{PLLeigen}
\end{equation}

\begin{equation}
    \mathcal{P}\sigma_+ \otimes \rho_B= \sigma_+\otimes \rho_B,
    \label{Peigen}
\end{equation}
where the eigenvalue
\begin{equation}
\begin{aligned}
    f(t-t_1) &\equiv \operatorname{Tr}_B[B(t)B^\dagger(t_1)\rho_B]e^{i\omega_0(t-t_1)} \\
    &= \int d\omega\, J(\omega) \exp[i(\omega_{0} - \omega)(t - t_{1})]
\end{aligned}
\end{equation}
is the bath correlation function, with $J(\omega)$ as the bath spectral density. \eqref{PLLeigen} and \eqref{Peigen} allow us to pass $\sigma_+ \otimes \rho_B$ through each term of the TCL-ME expansion, with the following eigenvalue (after taking the partial trace in order to get the reduced generator):

\begin{equation}
\begin{aligned}
\gamma_{2n}(t) &+ i S_{2n}(t) =
\int_{0}^{t} dt_{1} \int_{0}^{t_{1}} dt_{2} \cdots \int_{0}^{t_{2n-2}} dt_{2n-1} \\
&\times 2(-1)^{n+1} \langle f(t - t_{1}) f(t_{2} - t_{3}) \cdots f(t_{2n-2} - t_{2n-1}) \rangle_{OC}.
\end{aligned}
\end{equation}

For the adjoint terms, we recognize that $B\rho_B = 0$ in this model, allowing us to use 

\begin{equation}    \mathcal{P}^\dagger\mathcal{L}\mathcal{P} = \mathcal{P}\mathcal{L}\mathcal{P}^\dagger = 0
\end{equation}
as in Sec. IV C for this particular case. Symbolically expanding up to sixth order and canceling, we find only two nonzero terms contributed by the adjoint set. These are of the form

\begin{equation}
    \mathcal{P}\mathcal{L}\mathcal{L}\mathcal{P}\mathcal{P}^\dagger\mathcal{L}\mathcal{L}\mathcal{P}\mathcal{L}\mathcal{L}\mathcal{P},
    \label{newterm1}
\end{equation}
and 

\begin{equation}
    \mathcal{P}\mathcal{L}\mathcal{L}\mathcal{P}^\dagger\mathcal{L}\mathcal{L}\mathcal{P}\mathcal{L}\mathcal{L}\mathcal{P}.
    \label{newterm2}
\end{equation}

However, if we try to pass $\sigma_+ \otimes \rho_B$ through $\mathcal{P}\mathcal{P}^\dagger$ or $\mathcal{P}\mathcal{L}\mathcal{L}\mathcal{P}^\dagger$ , we find that although we preserve $\sigma_+ \otimes \rho_B$ as an eigenvalue, we end up taking the trace of the bath identity $I_B$. For continuous baths---the kind which the JC model is interested in---both \eqref{newterm1} and \eqref{newterm2} diverge.

Indeed, we can see this divergent behavior if we truncate the dimension of the bath and examine the behavior of $\rho_{11}(t)$ as we increase the cutoff. We take a classic example of a spectral density for which the TCL-ME breaks down \cite{Breuer1999StochasticWaveFunction}:

\begin{equation}
    J(\omega) = \frac{\gamma_{0}}{2\pi} \, \frac{1}{1 + \left[ \frac{\omega - \Omega_{0}}{\nu_{B}} \right]^{2}}.
\end{equation}

Using this Lorentzian spectral density, simulation results give Fig. 1 for different dimensions of the bath. It is clear how the divergent behavior emerges and worsens as we let the bath dimension increase.

\subsection{Ising Model}
Motivated by the failure of the TCL+-ME in the case of an infinite bath (as seen in the Jaynes-Cummings Model), we  turn our attention to the Ising model - replacing the harmonic oscillator bath with a discrete bath of qubits \cite{Krovi2007}. In this model, the interaction Hamiltonian $H_I$ is given by

\begin{equation}
    H_I = \sigma_z \otimes B,
\end{equation}
where the (shifted) bath operator is defined 

\begin{equation}
    B \equiv \sum_{n=1}^Ng_n\sigma_n^z - \theta I,
    \label{BIsing}
\end{equation}

\begin{equation}
    \theta \equiv Tr[\sum_{n=1}^Ng_n\sigma_n^z\rho_B].
    \label{theta}
\end{equation}

The reference state $\rho_B$ we take to be the thermal equilibrium ``Gibbs'' state:

\begin{equation}
    \rho_B \equiv \exp(-H_B/k_BT)/\operatorname{Tr}[\exp(-H_B/k_BT)].
\end{equation}

To apply the TCL in this case, we use the time-independence of $H_I$ to simplify $\mathcal{K}(t)$:

\begin{equation}
    \mathcal{K}(t) = \sum_{n=1}^{\infty} \lambda^n \frac{t^{\,n-1}}{(n-1)!} \langle \mathcal{L}^n \rangle_{oc}.
\end{equation}

As usual, to compute the TCL+ expansion, we need to add those combinations of $\mathcal{L}$s and $\mathcal{P}$s and their adjoints to the ordered cumulant $\langle \mathcal{L}^n \rangle_{oc}$ terms according to their power of $\lambda$.

To simplify the expansion of the commutators in each term, we first note that the Hamiltonian commutes with itself at all times, letting us use (for $n$ commutators)

\begin{equation}
   \underbrace{[H,[H,\cdots,[H,x]]]}_{n} = \sum_{k=0}^{n} (-1)^k \binom{n}{k} H^{\,n-k} \, \rho \, H^{\,k}.
\end{equation}

Further, letting $\rho = \rho_S \otimes \rho_B$, we note that 

\begin{equation}
\begin{aligned}
    &\operatorname{Tr}_B\bigg[\underbrace{[\sigma_z \otimes B,[\sigma_z \otimes B,\cdots,[\sigma_z \otimes B,\rho_s\otimes\rho_B]]]}_{n \,}\bigg] \\&\propto
    \begin{cases}
        i(\sigma_z\rho_S - \rho_S\sigma_z) & n \text{ odd} \\
        \sigma_z\rho_S\sigma_z - \rho_S & n \text{ even}.
    \end{cases}
    \label{prop}
\end{aligned}
\end{equation}

Because the $n$th term of the TCL will exclusively contain terms with $n$ $\mathcal{L}$s (since each $\mathcal{L}$ carries along with it a factor of $\lambda$), we can write  
\begin{equation}
\begin{aligned}
    \partial_t \rho_S = \mathcal{K}_S(t)\rho_S &= F_1(t)(\sigma_z\rho_S - \rho_S\sigma_z) \\
    &+ F_2(t)(\sigma_z\rho_S\sigma_z - \rho_S),
\end{aligned}
\end{equation}
where again we've used the reduced $\mathcal{K}_S(t)$ since we only care about the system dynamics, and $F_1(t)$ and $F_2(t)$ are constants of proportionality from \eqref{prop}. 

We can solve this equation in the Pauli basis to obtain:

\begin{align}
v_x(t) &= f(t)\big[v_x(0)\cos(g(t)) + v_y(0)\sin(g(t))\big], 
\\
v_y(t) &= f(t)\big[v_y(0)\cos(g(t)) - v_x(0)\sin(g(t))\big],
\\
v_z(t) &= v_z(0),
\end{align}

for the Bloch vector 
\begin{equation}
    \begin{bmatrix}
    v_x \\
    v_y \\
    v_z
    \end{bmatrix}
\end{equation}

with 
\begin{align}
    f(t) &= \exp\left(2\int_0^tF_1(t)dt\right),\\
    g(t) &= 2\int_0^tF_2(t)dt.
\end{align}

Armed now with the general solution for the TCL-n, which each new term alternately contributing to $f(t)$ and $g(t)$, we now compute both functions for TCL-5, as this is the first term at which the adjoint terms contribute.

\begin{align}
    f_5(t) &= \exp\!\left(-2Q_2 \alpha^{2} t^{2} + \frac{2Q_4 - 6Q_2^{2}}{3}\,\alpha^{4} t^{4}\right), \\
    g_5(t) &= \frac{4}{3} Q_3 \lambda^{3} t^{3} + \frac{2}{60} \lambda^{5} t^{5} \left(-16 Q_5 + 160 Q_2 Q_3\right)
\end{align}

where the $Q_n$ are the bath correlation functions of order $n$, or 
\begin{equation}
   Q_n = \mathrm{Tr}\!\left[ \underbrace{B \cdots B}_{n} \, \rho_B \right],
\end{equation}
which are computed explicitly in Ref. \cite{Krovi2007}. The TCL+-5 expansion yields one non-vanishing term,

\begin{equation}
    2\mathcal{P}\mathcal{L}\mathcal{P}^\dagger\mathcal{L}\mathcal{L}\mathcal{P}\mathcal{L}\mathcal{L}\mathcal{P}
\end{equation}
which, when applied on the state $\rho_S \otimes \rho_B$, yields 

\begin{equation}
    2\operatorname{Tr}[B](\operatorname{Tr}[B^2\rho_B^2] + \operatorname{Tr}[(B\rho_B)^2])(\rho_S \sigma_z - \sigma_z \rho_S)
    \label{isingplusterm}
\end{equation}
in keeping with our odd-even relation \eqref{prop} as expected. The traces in \eqref{isingplusterm} are computed explicitly in Appendix C.

For simulation, we test $N = 1$ and $N = 15$, as well as $\beta = 1$ and $\beta = 10$, and compare the results to those of the standard TCL-2, 4, and 5 (Fig. 2). It can be seen that the TCL+ as currently constructed results in a significant drop in accuracy, with this drop increasing significantly for higher $N$ and higher $\beta$ (lower temperature). Mathematically, this behavior results from the trace terms in \eqref{isingplusterm} introducing a factor of $2^N$, much like we had in the Jaynes-Cummings model (although the bath dimension is finite here so we get a convergent result), which forces the TCL+ prediction to blow up with a larger bath. Because the 5th order term only affects g(t), we only see this impact in the period of oscillation, but similar amplitude inaccuracies would follow were we to introduce the 6th-order term.

\section{SERIES CONVERGENCE ANALYSIS}
Because of the failure of the TCL+-ME to provide any sort of meaningful improvement to TCL breakdown as expected, we find it pertinent to analyze the convergence of Israel and Charnes' pseudoinverse expansion as described in Sec. II.

First, the well-known Neumann series for the inverse matrix 

\begin{equation}
(I - \Sigma)^{-1} = \sum_{k=0}^{\infty} \Sigma^k
\end{equation}
has the condition $\|\Sigma\| \leq 1$ to guarantee convergence, where we denote the operator norm by $\|\cdot\|$. Israel and Charnes' result \eqref{mpseries} for $A = I - \Sigma$ requires $\|I - \Sigma\| < \sqrt{2}$. These are fundamentally different conditions---if the operator norm of $\Sigma$ is $x$, then through the triangle inequality, all we can say about the latter is that 

\begin{equation}   
    |1 - x| \le \|I - \Sigma\| \le 1 + x.
\end{equation}

Recalling the definition of $\Sigma(t)$ from \eqref{Sigma}, we see from the integral bounds that at $t = t_0$, $\Sigma(t) = 0$. The classical breakdown of the TCL-ME happens when $I - \Sigma(t)$ is no longer invertible, i.e. $\Sigma(t)$ becomes too close to the identity. Therefore, the evolution of every case of breakdown in the TCL involves $\|\Sigma\|$ varying from 0 to 1 and beyond. 

The problem with the pseudoinverse series in this case is that as soon as $\|\Sigma\| \geq 1 - \sqrt{2} \approx 0.41$, the series no longer has a 100\% chance of converging. We can compare the two series' performance by measuring how fast they converge on the inverse of random (invertible) $I - \Sigma$ matrices with specified norms. Convergence for the geometric series as a function of series depth takes the form of exponential decay:

\begin{equation}
\|\sum_{k=0}^{d} \Sigma^k - [I - \Sigma]^{-1}\|
\sim \alpha e^{-d/\tau_i}, \\[6pt]
\end{equation}
\begin{equation}
\begin{aligned}
&\|\sum_{k=0}^{d} (\mathcal{I}-[\mathcal{I}-\Sigma(t)]^{\dagger}
   [\mathcal{I}-\Sigma(t)])^{k}[\mathcal{I}-\Sigma(t)]^{\dagger} 
   \\&- [I - \Sigma]^{-1}\|\sim \beta e^{-d/\tau_p},
\end{aligned}
\end{equation}
where the "depth constants" $\tau_i$ and $\tau_p$ describe the behavior of the convergence; a negative value indicates divergence. Plotting both depth constants as a function of matrix norm (Fig. 3a), we see that after the norm threshold of $\sqrt{2} - 1$, the pseudoinverse series very quickly stops converging. This is the reason for the failure of the TCL+: for the TCL to even travel into the breakdown regime, it must cross a norm threshold at which the series already diverges. Equivalently, no matter how well the TCL+ handles singular $I - \Sigma$, its significantly worse handling of standard invertible matrices compromise its viability.

This is not to say that there aren't cases of singular matrices which the pseudoinverse series treats better than the inverse, however. Take the matrix

\begin{equation}
    \Sigma =
    \begin{bmatrix}
    1 & 0 & 0 \\
    0 & 1.1 & 0 \\
    0 & 0 & 0.7
    \end{bmatrix}
    \label{exmat}
\end{equation}
for example. In this case $\|\Sigma\| = 1.1$, but $\|I - \Sigma\| = 0.3$, giving the advantage to the pseudoinverse series. Plotting norm error over depth reflects this (Fig. 3b).

\begin{figure}[t]
    \centering
    \begin{subfigure}[b]{\linewidth}
        \centering
        \includegraphics[width=\linewidth]{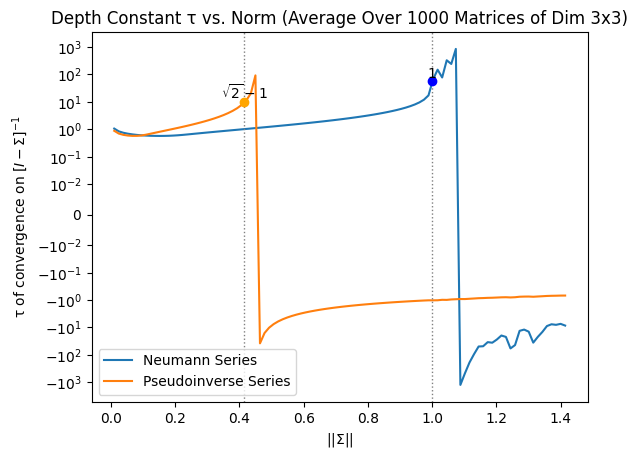}
        \caption{Depth constant produced by each series as a function of matrix norm.}
    \end{subfigure}
    \vfill
    \begin{subfigure}[b]{\linewidth}
        \centering
        \includegraphics[width=\linewidth]{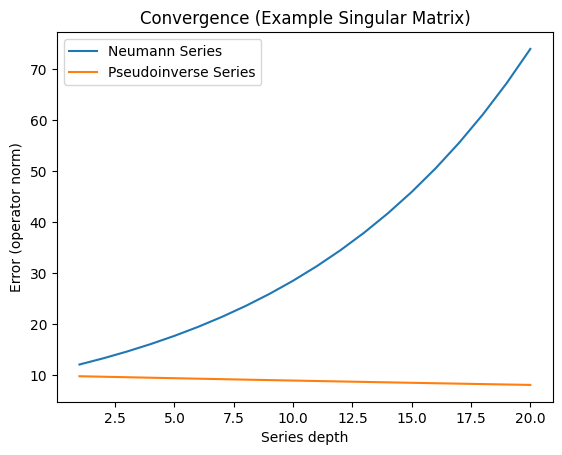}
        \caption{Convergence analysis of the matrix in \eqref{exmat}.}
    \end{subfigure}
    \caption{Depth constant and singular matrix convergence analyses to compare Neumann and pseudoinverse series.}
\end{figure}

Indeed, the pseudoinverse series will be convergent (and the Neumann series divergent) for any singular $\Sigma$ such that $\|I - \Sigma\|$ falls within $[0,\sqrt2)$. Thus, there may be some merit to a "concatenation" approach, in which we would "swap" master equations from the TCL to TCL+ as soon as the TCL breaks down. Of course, this approach would require knowing exactly when breakdown occurs, which can be difficult in general. Regardless, we leave this as the next avenue to be explored.

\section{CONCLUSION}
We find that the proposed improvement to the TCL-ME by modifying the operator $[I - \Sigma]^{-1}$ to use the more general Moore-Penrose inverse instead of the standard inverse is generally less accurate under perturbative expansion using Israel and Charnes' result, due to the expansion having a stricter convergence condition than the Neumann series. We outline certain cases where the expansion offers an improvement over the Neumann series and suggest a concatenation approach between both expansions, but the difficulty of predicting the moment of breakdown likely renders this method impractical for most problems. We hope that the search for a solution to TCL breakdown might continue in future research, and that perhaps the results demonstrated above provide some small help.

\section*{Acknowledgements}
I'd like to thank Micah Shaw for his help with simulation, as well as Dr. Daniel Lidar, Dr. Todd Brun, Juan Garcia Nila, and Joseph Barreto for their guidance throughout the course of the project.

\onecolumn
\appendix
\section*{APPENDIX}
\renewcommand{\thesection}{\Alph{section}.}
\renewcommand{\theequation}{\Alph{section}\arabic{equation}}
\setcounter{equation}{0}
\addcontentsline{toc}{section}{Appendix}
\section{PROOF FOR MOORE-PENROSE INVERSE SERIES EXPANSION}
Below we provide a proof for the power series expansion for the pseudoinverse described in Sec. II. Recalling \eqref{mpseries}:

\begin{equation}
A^{+}  = \sum_{k=0}^{\infty} (\operatorname{I}- A^{\dagger}A)^{k}A^{\dagger}
\end{equation}

By the singular value decomposition \( A = V D U^{\dagger} \):

\begin{equation}
\begin{aligned}
A^{k} &= \sum_{k=0}^{\infty} (\operatorname{I} - V D^{\dagger} U^{\dagger} U D V^{\dagger})^{k} V D^{\dagger} U^{\dagger} 
= \sum_{k=0}^{\infty} (V V^{\dagger} - V |D|^{2} V^{\dagger})^{k} V D^{\dagger} U^{\dagger} \\
&= \sum_{k=0}^{\infty} (V (I - |D|^{2}) V^{\dagger})^{k} V D^{\dagger} U^{\dagger} = \sum_{k=0}^{\infty} V (I - |D|^{2})^{k} D^{\dagger} U^{\dagger}.
\end{aligned}
\end{equation}

Now,
\begin{align}
\sum_{k=0}^{\infty} (\operatorname{I} - D^{\dagger} D)^{k} D^{\dagger}
&= \operatorname{diag} \left( \sum_{k=0}^{\infty} (1 - |d_i|^{2})^{k} d_i^{*} \right).
\end{align}

For \( |d_i|^{2} \le 2 \), this converges to
\begin{align}
\operatorname{diag}\!\begin{cases}
\frac{1}{d_i}, & d_i \neq 0, \\
0, & d_i = 0
\end{cases}
= \operatorname{diag}(d_i^{+}) = D^{+}
\end{align}

Adding back \( V \) and \( U^{\dagger} \):
\begin{align}
V D^{+} U^{\dagger} = A^{+}
\end{align}

Thus,
\begin{align}
A^{+} = \sum_{k=0}^{\infty} (I - A^{\dagger} A)^{k} A^{\dagger}
\end{align}

\section{CORRECTNESS PROOF FOR THE ADJOINTS OF $\mathcal{L}$ and $\mathcal{L}$}
\setcounter{equation}{0}
Here we prove the forms of $\mathcal{L}^{\dagger}$ and $\mathcal{P}^{\dagger}$ as defined in Sec. IV B (eqs. \eqref{Ldag} and \eqref{Pdag}). As a reminder, we defined the quantities: 
\begin{equation*}
    \mathcal{L} = -i[H(t),\cdot], \qquad \mathcal{L}^{\dagger} = i[H(t),\cdot] = \mathcal{L}^{*},
\end{equation*}

\begin{equation*}
    \mathcal{P}\rho = \operatorname{Tr}_{B}[\rho] \otimes \rho_{B}, \qquad
    \mathcal{P}^{\dagger} = \operatorname{Tr}_{B}[\cdot (I_{s} \otimes \rho_{b})] \otimes I_{b}.
\end{equation*}

$\mathcal{L}^{\dagger}$ follows trivially from linearity and the Hermiticity of $H(t)$; for $\mathcal{P}^{\dagger}$:

\begin{equation}
    \langle\mathcal{P}v,w\rangle = \langle v,\mathcal{P}^{\dagger}w\rangle.
\end{equation}

Using the Hilbert-Schmidt definition for the inner product of superoperators \eqref{schmidt}:

\begin{equation}
    \langle X,Y \rangle = \operatorname{Tr}[X^{\dagger}Y]
    \label{schmidt}
\end{equation}

\begin{equation}
    \operatorname{Tr}[(\mathcal{P}v)^{\dagger} w] = \operatorname{Tr}[v^{\dagger}(\mathcal{P}^{\dagger}w)]
\end{equation}

We may represent $v^{\dagger}$ and $w$ in terms of their respective system-bath basis vectors:

\begin{equation}
    v^{\dagger} = \sum_{i} A_{i}^{\dagger} \otimes B_{i}^{\dagger}, \qquad w = \sum_{j} C_{j} \otimes D_{j}.
\end{equation}

Where $A_{i}, C_{j} \in \mathcal{H}_{S}$ and $B_{i}, D_{j} \in \mathcal{H}_{B}$.

For the left side, the inner product simplifies to

\begin{equation}
\begin{aligned}
\operatorname{Tr}[\operatorname{Tr}_{B}[v^{\dagger}] \otimes \rho_{B} w] =
\sum_{i,j} \operatorname{Tr}[B_{i}]^{*}\operatorname{Tr}[A_{i}^{\dagger} C_{j}]\operatorname{Tr}[D_{j}\rho_{B}].
\label{Pleft}
\end{aligned}
\end{equation}

For the right  side, we have (using the definition of $\mathcal{P}^{\dagger}$ defined in \eqref{Pdag}:

\begin{equation}
\begin{aligned}
    \langle v, \operatorname{Tr}_{B}[w(I_{S} \otimes\rho_{B})] \otimes I_{B} \rangle = \operatorname{Tr}[v^{\dagger} \operatorname{Tr}_{B}[w(I_{S} \otimes\rho_{B})] \otimes I_{B}],
    \end{aligned}
\end{equation}
which simplifies to

\begin{equation}
    \sum_{i,j} \operatorname{Tr}[B_{i}]^{*}\operatorname{Tr}[A_{i}^{\dagger}C_{ j}]\operatorname{Tr}[D_{j}\rho_{B}].
    \label{Pright}
\end{equation}

That \eqref{Pright} matches \eqref{Pleft} validates our definition of $\mathcal{P}^{\dagger}$ \eqref{Pdag}.

\section{ISING MODEL TRACE QUANTITIES}
\setcounter{equation}{0}
Here we calculate the quantities $\operatorname{Tr}[B]$,  $\operatorname{Tr}[B^2\rho_B^2]$, and $\operatorname{Tr}[(B\rho_B)^2]$, which are present in the extra TCL+ term for the Ising model \eqref{isingplusterm}.

Recalling \eqref{BIsing} and \eqref{theta}:

\begin{equation}
    B \equiv \sum_{n=1}^Ng_n\sigma_n^z - \theta I,
    \label{BIsing}
\end{equation}

\begin{equation}
    \theta \equiv \operatorname{Tr}[\sum_{n=1}^Ng_n\sigma_n^z\rho_B].
    \label{theta}
\end{equation}

We may write the Gibbs state $\rho_B$ as 

\begin{equation}
\rho_B = \bigotimes_{n=1}^{N} 
\frac{\exp\left(-\frac{\Omega_n}{2kT} \sigma_n^z \right)}
{\mathrm{Tr}\left[\exp\left(-\frac{\Omega_n}{2kT} \sigma_n^z \right)\right]}
=
\bigotimes_{n=1}^{N} \frac{1}{2} \left(I + \beta_n \sigma_n^z \right)
\equiv \prod_{n=1}^{N} \rho_n,
\end{equation}

where 
\begin{equation}
\beta_n = \tanh\!\left( - \frac{\Omega_n}{2 k T} \right).
\end{equation}

Using this definition, we can simplify $\theta$:

\begin{equation}
\begin{aligned}
\theta = \mathrm{Tr} \Bigg[ \sum_{n=1}^{N} g_n \sigma_n^z \bigotimes_{m=1}^{N} \frac{1}{2} \big(I + \beta_m \sigma_m^z \big) \Bigg] 
= \sum_{n=1}^{N} g_n \, \mathrm{Tr} \Big[ \frac{1}{2} (\sigma_n^z + \beta_n I) \Big] \prod_{m \neq n} \mathrm{Tr} \Big[ \frac{1}{2} (I + \beta_m \sigma_m^z) \Big] 
= \sum_{n=1}^{N} g_n \beta_n.
\end{aligned}
\end{equation}

For $\operatorname{Tr}[B]$,
\begin{equation}
\begin{aligned}
\operatorname{Tr}[B] = \operatorname{Tr}[\sum_{n=1}^Ng_n\sigma_n^z - \theta I] = \sum_{n=1}^Ng_n\operatorname{Tr}[\sigma_n^z] - \sum_{n=1}^{N} g_n \beta_n\operatorname{Tr}[I] = -2^N  \sum_{n=1}^{N} g_n \beta_n.
\end{aligned}
\end{equation}

For $\operatorname{Tr}[B^2\rho_B^2]$,
\begin{equation}
\begin{aligned}
\rho_B^2 = \bigotimes_{n=1}^N \frac{1}{4}(I + \beta_m\sigma_m^z)^2 = \bigotimes_{n=1}^N\frac{1}{4}[(1 + \beta_n^2)I + 2\beta_n\sigma_n^z]
\end{aligned}
\end{equation}

\begin{equation}
\begin{aligned}
B^2 &= \sum_{m,n=1}^N g_m g_n \sigma_m^{z} \sigma_n^{z} 
    - 2\Theta \sum_{n=1}^N g_n \sigma_n^{z} 
    + \Theta^2 I
\end{aligned}
\end{equation}

\begin{equation}
\begin{aligned}
\operatorname{Tr}(B^2 \rho_B^2) 
&= \frac{1}{4} \sum_{m,n=1}^N g_m g_n 
   \operatorname{Tr}\!\left[ 
   (\sigma_m^{z} \sigma_n^{z}) 
   \!\bigotimes_{\ell=1}^N 
   \!\!\left((1+\beta_\ell^2) I 
   + 2\beta_\ell \sigma_\ell^{z} \right)
   \right] 
 - \frac{2\Theta}{4} 
   \sum_{n=1}^N g_n 
   \operatorname{Tr}\!\left[
   \sigma_n^{z} \!\bigotimes_{\ell=1}^N 
   \!\!\left((1+\beta_\ell^2) I 
   + 2\beta_\ell \sigma_\ell^{z} \right)
   \right] \\[6pt]
& + \frac{\Theta^2}{4}
   \operatorname{Tr}\!\left[
   \bigotimes_{\ell=1}^N
   \!\left((1+\beta_\ell^2) I 
   + 2\beta_\ell \sigma_\ell^{z} \right)
   \right]
\end{aligned}
\end{equation}

\begin{equation}
\begin{aligned}
&= \frac{1}{4} \sum_{m\neq n} g_m g_n 
   \operatorname{Tr}\!\left[
   \big((1+\beta_m^2)\sigma_m^{z}
   + 2\beta_m I\big)
   \right]\operatorname{Tr}\!\left[
   \big((1+\beta_n^2)\sigma_n^{z}
   + 2\beta_n I\big)
   \right]
   \prod_{\ell\neq m,n} \operatorname{Tr}\!\left[
   \big((1+\beta_\ell^2)\sigma_\ell^{z}
   + 2\beta_\ell I\big)
   \right] \\[6pt]
& + \frac{1}{4} \sum_{m=n} g_m^2 
   \prod_{\ell\neq m} \operatorname{Tr}\!\left[
\big((1+\beta_\ell^2)\sigma_\ell^{z}
   + 2\beta_\ell I\big)
   \right] + \frac{\Theta}{2}\sum_{n=1}^N g_n^2
   \operatorname{Tr}\!\left[
   ((1+\beta_n^2)\sigma_n^{z}
   + 2\beta_n I)
   \right]
   \prod_{\ell\neq n}^N
   \operatorname{Tr}\!\left[
   ((1+\beta_\ell^2) I
   + 2\beta_\ell \sigma_\ell^{z})
   \right] \\[6pt]
&+ \frac{1}{4}\Theta^2
   \prod_{\ell=1}^N
   \operatorname{Tr}\!\left[
   ((1+\beta_\ell^2) I
   + 2\beta_\ell \sigma_\ell^{z})
   \right]
\end{aligned}
\end{equation}

\begin{equation}
\begin{aligned}
&= 2^N \sum_{m\neq n} g_m g_n \beta_m \beta_n
   \prod_{\ell\neq m,n} (1+\beta_\ell^2)
   + 2^N \sum_{m=n} g_m^2 
   \prod_{\ell=1}^N (1+\beta_\ell^2) - 2^N \Theta \sum_{n=1}^N 
   g_n \beta_n \prod_{\ell\neq n}
   (1+\beta_\ell^2)
   + 2^{N-2}\Theta^2 \prod_{\ell=1}^{N}
   (1+\beta_\ell^2)
\end{aligned}
\end{equation}

Because $B$ and $\rho_B$ commute, the result is the same for $\operatorname{Tr}[(B\rho_B)^2]$.
\twocolumn
\bibliographystyle{unsrt}
\bibliography{references}
\end{document}